# GREEN WSN- OPTIMIZATION OF ENERGY USE THROUGH REDUCTION IN COMMUNICATION WORKLOAD

Vandana Jindal[1], A.K.Verma[2] and Seema Bawa[2]
[1,2]Department of Computer Science and Engineering, Thapar University, Patiala, India
jindal_vandana@yahoo.co.in

## ABSTRACT

*Applications of Wireless Sensor Networks (WSNs) are growing day by day due to the ease of deployment, reduction in costs to affordable levels and versatility of these networks. Besides developing advanced micro fabrication technologies means are being devised to reduce energy consumption to bring the network setup and operational costs down. With increasing applications amount of energy consumed in these networks is enormous. Even a small savings in energy consumption will result in huge benefits in energy consumption globally. Bulk of the energy is consumed in communication activity of these networks. It is our endeavour to optimize this activity to make these networks energy efficient and thereby reducing their impact on the overall environment in line with the principle 'Go Green'.*

## 1. INTRODUCTION

With advances in micro fabrication technologies there has been an exponential growth in the WSN applications, as the nodes are becoming cheaper and more versatile. Till now, these networks were considered for high tech and higher end applications only as the costs involved were very high. Though WSNs are finding use in every walk of life however, all pervasive use of these networks find major deterrent in limited battery life of the sensor nodes which are the building blocks of these networks. This limitation has shifted the focus of research community towards devising ways and means for reducing energy consumption which will increase the life of the network and at the same time the applications will become more 'Green' or eco friendly. Research activities are being carried out in various directions to achieve this objective. Major efforts are towards devising specific operating systems, optimizing sensor awake and sleep schedules, designing middleware using WSN specific query optimization and data processing protocols. Major chunk of energy is consumed in data communication therefore efforts to reduce communication traffic are going to bring about maximum benefits. Multi hop communication is one of the first steps in this direction. Communication payload can be optimized by In-network data processing and query optimization at the sink node. Useful In-network processing may employ one or all of the processes such as data compilation, data compression, data aggregation etc. Low power compilation which is not much talked of in case of sensor networks has a potential to produce substantial reductions in energy requirements. Data aggregation enroute to the sink node is another effective means to contain energy costs by aggregation of communication pay loads. Besides these In-network processes, query optimization at the base station reduces the number of queries injected into the network as results for new queries are searched in the existing data received from the nodes and the number of queries are merged to produce synthetic queries.

In this paper, we propose a framework which has been designed with a view to achieving reduction in communication costs by using in-network processing, base-station query optimization and modification of query syntax. Sec II deals with the related works followed by Sec III giving the architecture of the proposed framework. Sec IV shows the performance analysis followed by the conclusion in Sec V.

## 2. RELATED WORK

A lot of work has been done to study the effects of In-network processing such as re scheduling of instructions, optimization of compilation process and data aggregation at nodes. Works on Digital Signal Processors (DSPs), used for mobiles & embedded computing have been carried out by V.Tiwari et al. [1]. Reduction in energy consumption by 30% was observed by changing the sequence of the input operands while executing multiple instructions. Nicholas D.Lane et al. [2] conducted studies for examining the relationship between the instruction types, instruction operand order, memory addressing mode and energy consumption. A direct relationship between the execution of computational workloads and the associated energy consumption was observed.

Compiler based optimization is being used and a number of techniques [3-5] have been developed which can be adapted according to the specificity of the system where compilation is to be done. Off late there has been focus on data compressions in sensor nodes of WSNs to achieve energy savings.
Pradhan et al. [6] used an idea of distributed compression where source and channel coding were used resulting in data reduction between the nodes. Rabat et al. [7] proposed a system in which sensor data is predicted randomly and the data is broadcast throughout the network by gossip algorithm. Various subsets of nodes are framed on the basis of similar statistical data so that the data can be retrieved from a particular subset. Data approximation can be arrived at by reconstruction of the data.

In [8, 9] audio and video compressions have been discussed. Use of compression in routing, aggregation, indexing and storage, energy balancing is shown in [10].
Routing protocols and WSN architectures [11] were developed to send data through pre defined routes in first Generation of WSNs. Various MAC energy efficient protocols [12] along with In-network data aggregation [13] techniques were used. In second Generation, WSN was considered as a distributed data centric database [14, 15] where user can interact with this database with SQL queries and the focus was on devising energy efficient ways of executing such queries. In the third Generation [16] methods were adopted for multi query optimization and aggregation. Recently the fourth Generation of WSNs has witnessed developments in energy efficiency and storage capacity of local flash storage.
The work proposed by us uses In-network aggregation, base-station multi query optimization and modification of query syntax by using a variable in place of words used repeatedly in various queries. There has been no compromise on data integrity as no data approximations have been used. Operation of a multiuser WSN serving needs of multiple users for multiple applications has been optimized in this framework.

## 3. PROPOSED FRAMEWORK

A WSN is a distributed database where the data is generated continuously by sensing activity of sensors to serve various applications for which the WSN has been deployed. This database is to be handled judiciously so that required information can be extracted from it by using resources such as energy and time etc. optimally. All efforts by research community are towards achieving this goal. Our framework is also an endeavor in this direction.

To serve an application three processes i.e. sensing, data processing and transmission are required to be carried out in a WSN. All these three processes consume energy though to a varying extent. It is well known that bulk of the energy consumption is in the transmission phase of WSN data.

Studies have shown that 0.4µJ is the energy required to transmit a single bit of data whereas in compression of a single bit uses 0.86nJ only i.e. transmission is 480 times more energy consuming. Data compression techniques can be made more efficient by using low complexity small size data compression algorithms.

In response to each query, the data sensed by the motes has to be transmitted towards the base station. In order to decrease the energy consumption in the WSN, emphasis has been laid on the reduction in number and size of the data transmissions.

In-network compression and modification in Query syntax help in reducing the size of data to be transmitted from node to node and nodes to base station and Query optimization at the base-station (sink node) reduces the number of queries injected into the network and thereby reduces the transmissions in response to queries.

In the latest WSNs, advanced In-network processing is possible due to increase in the computing capacity of sensor nodes. The method proposed and implemented by us uses both the above mentioned optimization techniques. There are two stages in our optimization process. In Stage-I, the data stream is subjected to compression in the network and in Stage-II Query merging and use of variable in query syntax is employed at base-station.

The framework is composed of the following components:

*User:* A person desiring to retrieve information sitting in the lab/ home/ workplace does this by injecting a query or number of queries from his PC. The query he injects from his PC is sent via the base-station to the free nodes (motes).

*Motes:* Sun Microsystems have developed a mote SunSPOT which is built upon the IEEE 802.15.4 standard. The mote is built on Squawk Virtual Machine (VM) [59]. Size of a SPOT is nearly equal to the size of a 3×5 card. It consists of 32-bit ARM9 CPU, 1 MB RAM and 8 MB of Flash memory, a 2.4 GHz radio and a USB interface. The network platform of SPOT has built-in sensors along with the capability of interfacing with external devices. Two kinds of SPOTs i.e. free-range SPOT and base-station SPOT are present. The anatomy of free range SPOT has a battery processor board, a sensor board and a sunroof.

*Base-station:* The base-station SPOT differs from the WSN free nodes in terms of absence of sensor board. It acts as an interface between the base station application running on the host (PC with Windows platform) and those running on the targets. Energy source available at the base-station is replenishable.

In response to a query injected at the base-station the processes involved in the transmission of the sensed data from the motes to the base station are:

*In-network optimization*: The input of the motes i.e. the sensed data is in the form of continuous data stream. This continuous data stream is tokenized, generating strings of data. These generated strings are subjected to various compression algorithms. Algorithms used are Huffman, LZW and Deflate. Compression results of these algorithms were studied and it was observed that Huffman compression is the best of these three.

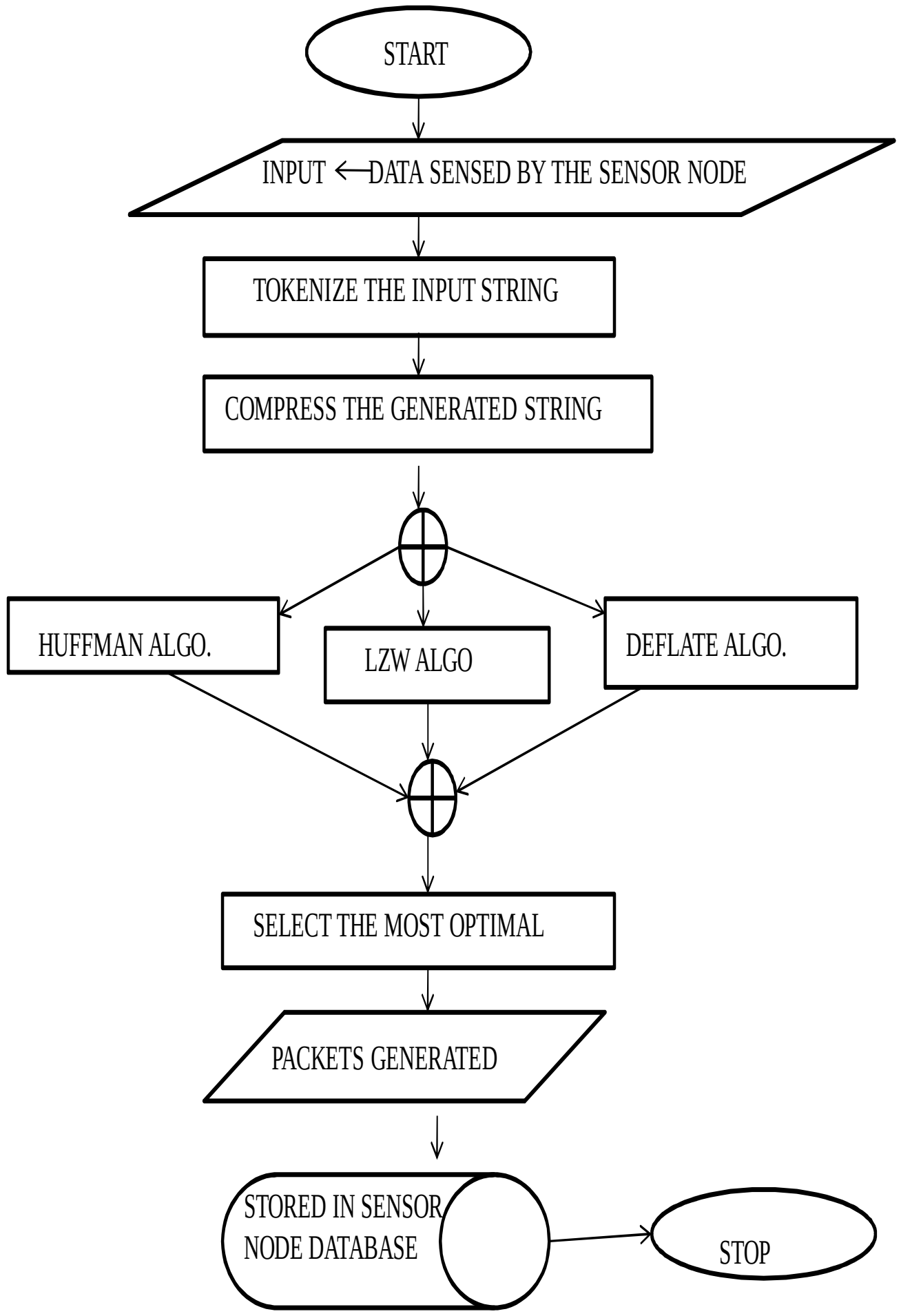

Figure 1: Flowchart showing In-network processing

*Base-station optimization*: At the base-station, query optimization is done using query merging technique. The optimization of multiple queries at the base-station helps in significant reduction in energy consumption. As the base-station is not resource constrained, it filters out the redundant load of multiple queries into the network. If a new query $q_n$ arrives at a base-station where result of a query set $Q_s$ have already been calculated or obtained from the network. The new query is merged with the existing queries and a synthetic query emerges out of this merger.

Algorithm has been devised which compares energy benefits by using new query and already existing queries as it is and synthetic query. A '*Gain*' metric has been developed to compare the Gain of query merging. In case, Gain is positive synthetic query is used. Slowly existing query set transforms into set of synthetic queries. Once requirement of any query is fulfilled it is removed from the query set. In addition to this as the queries injected into the network have identical expressions to a large extent these identical expressions are replaced with single characters called "*Static Variables*" which further compresses the data to be transmitted in the network.

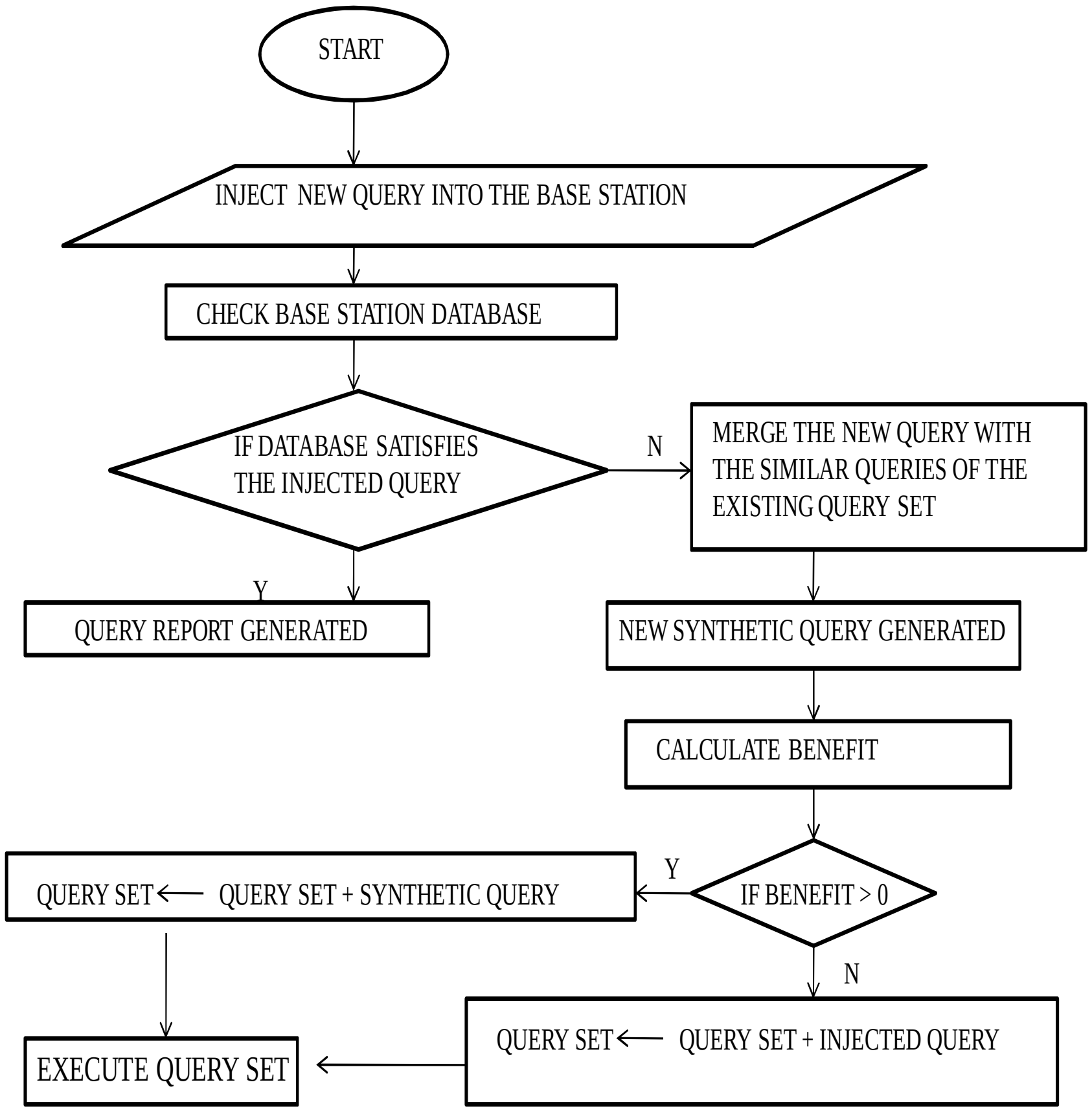

Figure 2: Flowchart showing Base station optimization

The proposed framework "**C***ompression* **A***t inpu***T** *with* **M***ulti-query* **O***ptimization at base-***S***tation"* **(CATMOS),** runs on Java VM platform. In this work we depict how CATMOS compresses the query syntax and the sensed data along with query merging at the base-station before presenting the final results in response to the user's queries.

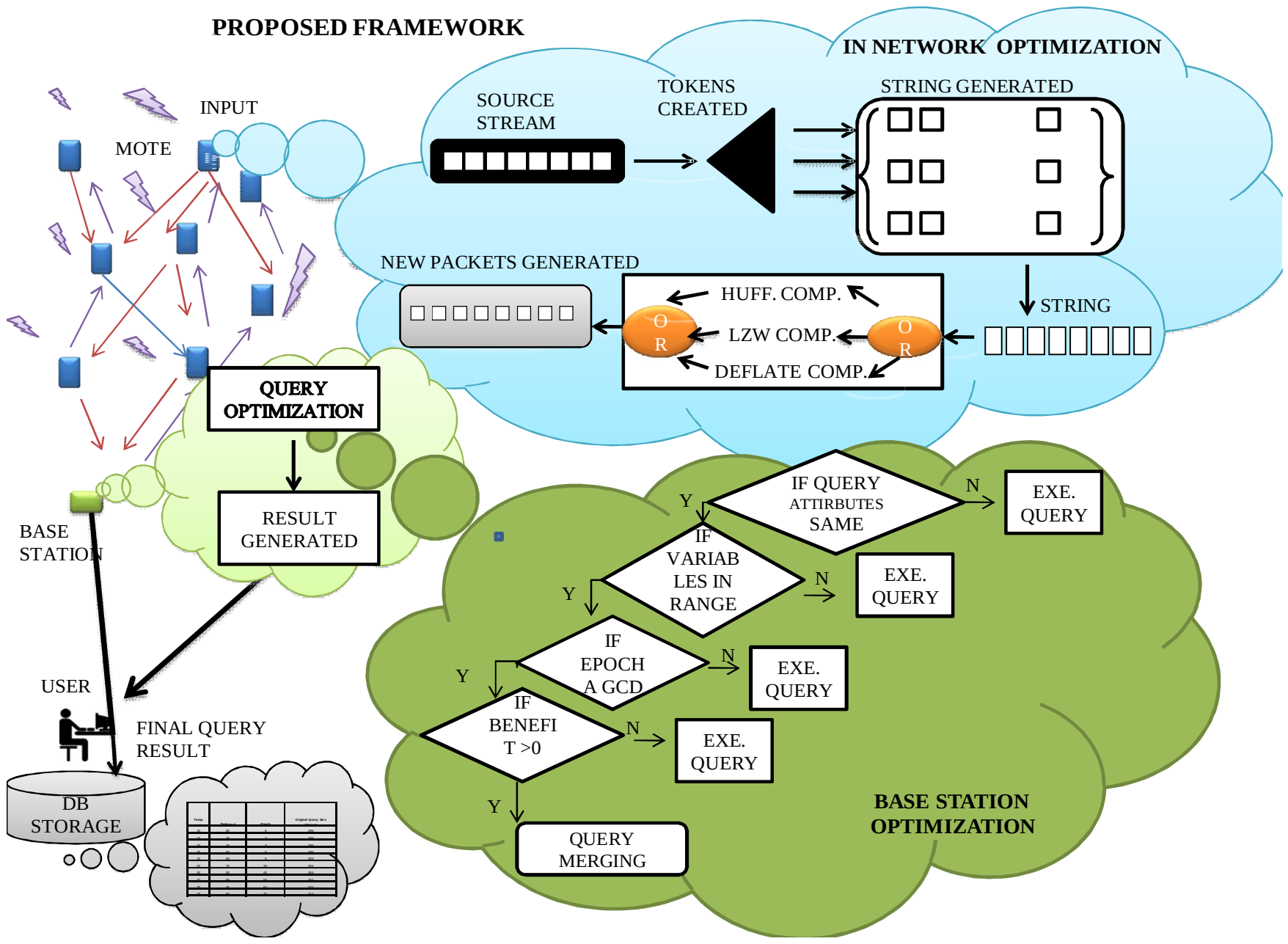

Figure3: The proposed architecture.

## 4. Performance Analysis

The findings have been summarized in tabular form and then the graphical representations have been depicted.

Table 1:Compression table showing the original query size (in bytes) along with the Query size after compression using Huffman's/ LZW/ Deflate algorithm.

| **$Temp_{(min)}$** | **$Temp_{(max)}$** | **Epoch** | **Original Query size (bytes)** | **Query size using Huffman Algo. (bytes)** | **Query size using LZW Algo. (bytes)** | **Query size using Deflate Algo. (bytes)** |
|---|---|---|---|---|---|---|
| 10 | 50 | 5 | 296 | 152 | 282 | 360 |
| 10 | 40 | 8 | 296 | 144 | 282 | 360 |
| 10 | 50 | 5 | 304 | 152 | 290 | 368 |
| 10 | 35 | 40 | 304 | 152 | 290 | 368 |
| 15 | 25 | 50 | 304 | 152 | 290 | 368 |
| 15 | 30 | 10 | 304 | 152 | 290 | 352 |
| 5 | 55 | 120 | 312 | 160 | 305 | 376 |
| 25 | 50 | 130 | 312 | 160 | 291 | 376 |
| 30 | 45 | 60 | 304 | 152 | 297 | 368 |
| 35 | 50 | 70 | 312 | 160 | 298 | 376 |

Table 2: Table showing the Compression Factor(%age compression using Huffman's/ LZW/ Deflate algorithm).

| **Huffman (Compression Factor)** | **LZW (Compression Factor)** | **Deflate (Compression Factor)** |
|---|---|---|
| 51.351351 | 95.27027 | 121.62162 |
| 48.648649 | 95.27027 | 121.62162 |
| 50 | 95.394737 | 121.05263 |
| 50 | 95.394737 | 121.05263 |
| 50 | 95.394737 | 121.05263 |
| 50 | 95.394737 | 121.05263 |
| 50 | 95.394737 | 115.78947 |
| 51.282051 | 97.75641 | 120.51282 |
| 50 | 97.697368 | 121.05263 |
| 51.282051 | 95.512821 | 120.51282 |

Table 3: Difference in amount of %age compression (using Huffman / LZW Compression algorithm).

| **Huffman (Comp. Factor)** | **Huffman (Comp. Factor with Static Var.)** | **Huffman (Comp. Diff)** | **LZW algorithm (Comp. Factor)** | **LZW algorithm (Comp. Factor with Static Var.)** | **LZW algorithm (Comp. Diff)** |
|---|---|---|---|---|---|
| 51.351351 | 42.105263 | 9.246088 | 95.270270 | 95.394736 | -0.124466 |
| 48.648648 | 36.842105 | 11.806543 | 95.270270 | 90.789473 | 4.480796 |
| 50.000000 | 40.000000 | 10.000000 | 95.394736 | 95.625000 | -0.230263 |
| 50.000000 | 40.000000 | 10.000000 | 95.394736 | 95.625000 | -0.230263 |
| 50.000000 | 40.000000 | 10.000000 | 95.394736 | 95.625000 | -0.230263 |
| 50.000000 | 40.000000 | 10.000000 | 95.394736 | 91.250000 | 4.144736 |
| 51.282051 | 42.857142 | 8.424908 | 97.756410 | 100.000000 | -2.243589 |
| 51.282051 | 42.857142 | 8.424908 | 93.269230 | 91.666666 | 1.602564 |
| 50.000000 | 40.000000 | 10.000000 | 97.697368 | 100.000000 | -2.302631 |
| 51.282051 | 38.095238 | 13.186813 | 95.512820 | 95.833333 | -0.320512 |

Results of the study are shown in the graphs. Results of compression of sensed data with three algorithms: Huffman, LZW and Deflate are shown in Figures 4a, 4b, 4c respectively.

As is very evident Graphs 4(a) and 4(b) show the results as expected i.e. we see the reduced size of the queries after compressing the queries using Huffman code and LZW algorithm. The third graph i.e. 4(c) shows the results opposite to our expectations.

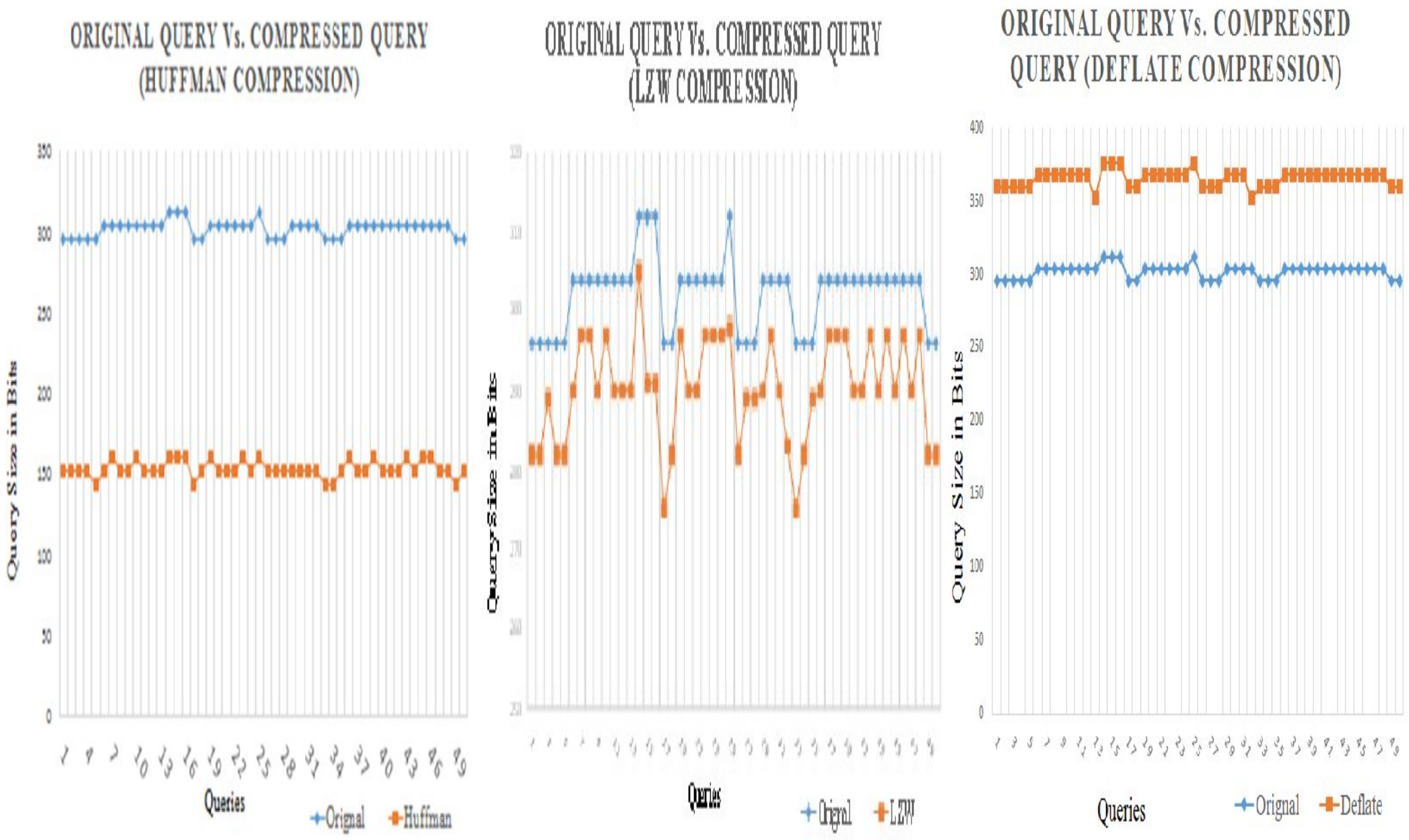

Figure.4: Graphs showing results of Original Query Vs. Compressed Query with
(a) Huffman (b) LZW(c) Deflate

Usage of static variables in query compression (with Huffman code and LZW algorithm), further improved the compression.

Because of the above results obtained (refer Graph 4 (c)), the Deflate algorithm is not considered for further compression using Static Variables. Results are shown in Figures 5a, 5b.

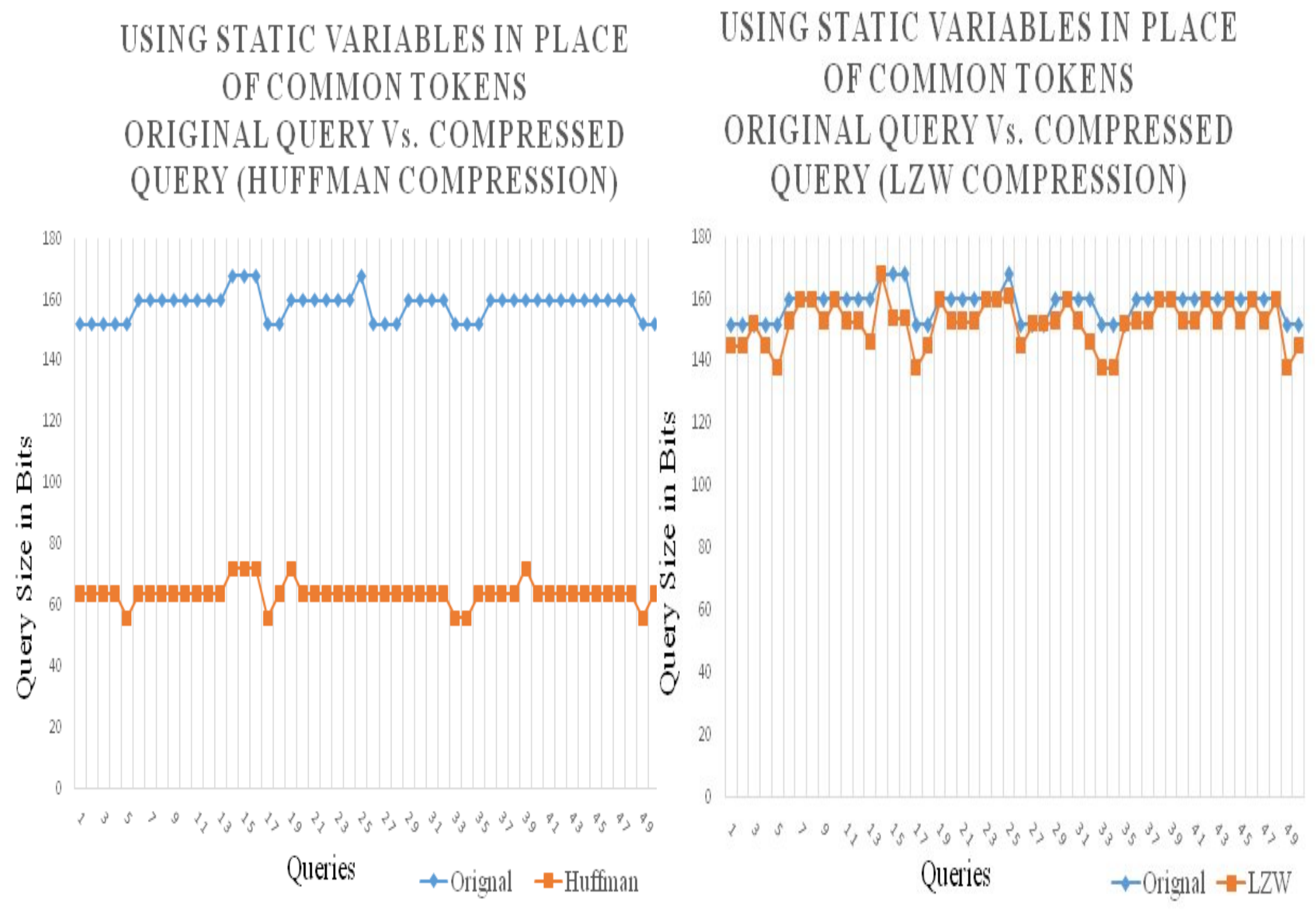

Figure.5: Graphs showing results of Compressed Query Vs. Compressed Query (Static Vars.) (a) Huffman (b) LZW

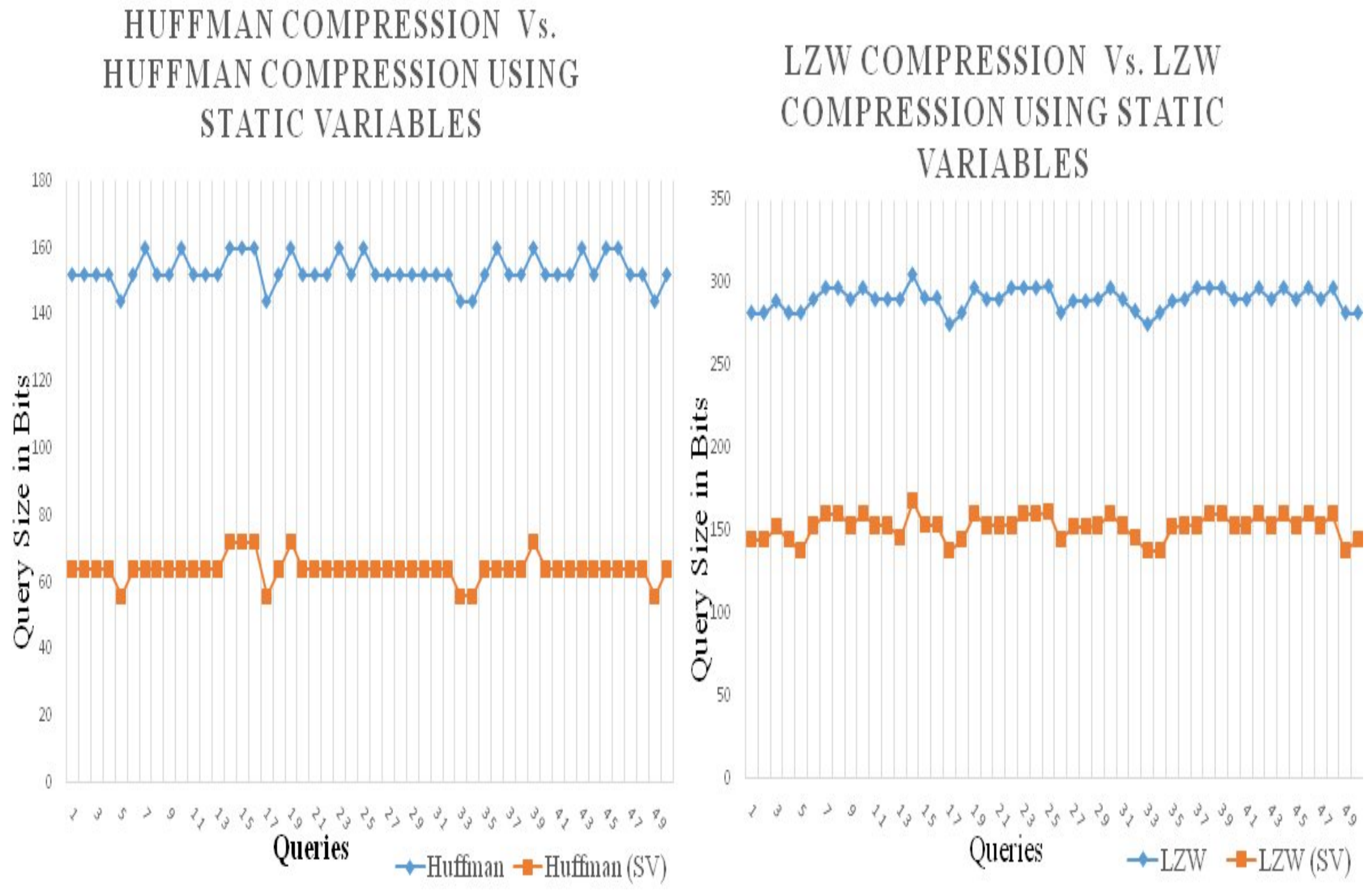

Figure.6: Graphs showing results of Improved energy gains (a) Huffman (b) LZW

The graphs depict clearly that queries transmitted after compression and the queries transmitted using static variables during program execution along with the compression prove to be a better choice as we achieve remarkable compression. The test was successfully conducted with queries when passed through three different compression algorithms i.e. Huffman compression, LZW and Deflate algorithm.

## 5. CONCLUSION AND FUTURE SCOPE

In this paper, we have proposed use of low power compilation techniques in WSNs and query merging for optimization at base station in order to decrease the energy consumption in a WSN. The application of this optimization resulted in a modest 10.29% reduction in the power consumption of the application. Energy savings in one network may seem to be miniscule but as the number of deployed WSNs is increasing day by day, these savings translate into huge amounts globally. This saving in energy will definitely reduce environmental effects and will help in achieving '*Green World*'. In addition to this, WSN can be deployed in any inaccessible location to monitor impending climatic or environmental hazards which can be contained if detected timely such as movements of glaciers, forest fires etc.

In short, reduction in energy consumption and possibility of deployment of WSN in any unimaginable location will go a long way in achieving environmental protection targets/ goals.

**Authors**

**Vandana Jindal** is currently working as an Assistant Professor in the department of Computer Science at D.A.V College, Bathinda. She holds degrees of B.Tech, MCA, M.Phil. Since January 2009, she has been with the Thapar University, Patiala in Punjab as a Ph.D. student. Her research interests include database management systems, wireless sensor networks. She is a member of IEEE and IEI.

**A. K. Verma** is currently working as Associate Professor in the department of Computer Science and Engineering at Thapar University, Patiala in Punjab (INDIA). He received his B.S. and M.S. in 1991 and 2001respectively, majoring in Computer Science and Engineering. He has worked as Lecturer at M.M.M. Engineering College, Gorakhpur from 1991 to 1996. His research interests include wireless networks, routing algorithms and securing ad hoc networks.

**Seema Bawa** holds M.Tech (Computer Science) degree from IIT Kharagpur and Ph.D. from Thapar Institute of Engineering & Technology, Patiala. She is currently Professor in the department of Computer Science and Engineering at Thapar University, Patiala in Punjab (INDIA). Her areas of interest include Parallel and distributed computing, Grid computing, VLSI Testing and network management. Prof. Bawa is member of IEEE, ACM, Computer society of India and VLSI Society of India.